\def\year{2022}\relax
\documentclass[letterpaper]{article} % DO NOT CHANGE THIS
\usepackage{aaai22}  % DO NOT CHANGE THIS
\usepackage{times}  % DO NOT CHANGE THIS
\usepackage{helvet}  % DO NOT CHANGE THIS
\usepackage{courier}  % DO NOT CHANGE THIS
\usepackage[hyphens]{url}  % DO NOT CHANGE THIS
\usepackage{longtable}
\usepackage{graphicx} % DO NOT CHANGE THIS
\def\UrlFont{\rm}  % DO NOT CHANGE THIS
\usepackage{natbib}  % DO NOT CHANGE THIS AND DO NOT ADD ANY OPTIONS TO IT
\usepackage{caption} % DO NOT CHANGE THIS AND DO NOT ADD ANY OPTIONS TO IT
\usepackage{pgf}
\usepackage{pgfplots}
\DeclareCaptionStyle{ruled}{labelfont=normalfont,labelsep=colon,strut=off} % DO NOT CHANGE THIS
\usepackage{algorithm}
\usepackage{enumitem}
\usepackage[]{hyperref}
\hypersetup{
    colorlinks=true,
    linkcolor=blue,
    citecolor=blue,
    filecolor=blue,      
    urlcolor=blue,
}

\usetikzlibrary{pgfplots.groupplots}
\usetikzlibrary{shapes.geometric}
\usetikzlibrary{positioning,fit,shapes.geometric,backgrounds, calc}
\usetikzlibrary{arrows,decorations.markings}
\usetikzlibrary{shapes.arrows}
\usetikzlibrary{patterns}
\tikzset{textnode/.style={inner sep=0pt,outer sep=0,execute at begin node={\strut}}}
\tikzstyle{state} = [textnode,circle, draw, inner sep=0pt, outer sep=0]
\usepgfplotslibrary{groupplots}

\pgfplotsset{scaled y ticks=false}
\pgfplotsset{scaled x ticks=false}
                    
\pgfplotsset{every axis/.append style={
                    xlabel={$x$},          % default put x on x-axis
                    ylabel={$y$},          % default put y on y-axis
                    label style={font=\sffamily\small},
                    tick label style={font=\sffamily\small},
                    xticklabel style = {font=\sffamily\scriptsize},
                    title style = {font=\footnotesize\sffamily},
                    ylabel near ticks,
                    y label style={font=\sffamily\scriptsize},
                    xlabel near ticks,
                    x label style={font=\sffamily\scriptsize},
                    legend cell align={left},
                    legend style={draw=none, font=\sffamily\scriptsize},
                    },
                    legend image code/.code={
                    \draw[mark repeat=2,mark phase=2]
                        plot coordinates {
                        (0cm,0cm)
                        (0.15cm,0cm)        %% default is (0.3cm,0cm)
                        (0.3cm,0cm)         %% default is (0.6cm,0cm)
                        };%
                    }
                    }
\pgfplotsset{compat=newest}

\usepackage{newfloat}
\usepackage{listings}
\floatstyle{ruled}
\newfloat{listing}{tb}{lst}{}
\floatname{listing}{Listing}
\usepackage{booktabs}

\title{Truth Social Dataset}
\author{
    Patrick Gerard, Nicholas Botzer, Tim Weninger
}
\affiliations{
    Department of Computer Science and Engineering\\
    University of Notre Dame\\
    Notre Dame, Indiana, USA\\
    \{pgerard2, nbotzer, tweninger\}@nd.edu
}

\begin{document}

\maketitle

\begin{abstract}
Formally announced to the public following former President Donald Trump’s bans and suspensions from mainstream social networks in early 2022 after his role in the January 6 Capitol Riots, Truth Social was launched as an ``alternative'' social media platform that claims to be a refuge for free speech, offering a platform for those disaffected by the content moderation policies of the existing, mainstream social networks. The subsequent rise of Truth Social has been driven largely by hard-line supporters of the former president as well as those affected by the content moderation of other social networks. These distinct qualities combined with its status as the main mouthpiece of the former president positions Truth Social as a particularly influential social media platform and give rise to several research questions. However, outside of a handful of news reports, little is known about the new social media platform partially due to a lack of well-curated data. In the current work, we describe a dataset of over 823,000 posts to Truth Social and and social network with over 454,000 distinct users. In addition to the dataset itself, we also present some basic analysis of its content, certain temporal features, and its network.
\end{abstract}

\section{Introduction}
\label{sec:introduction}
The social media platform \textit{Truth Social} was launched in February of 2022 about a year after the suspension of former United States President Donald Trump from Twitter, Facebook, and other social media platforms. Truth Social is largely stylized after Twitter where Tweets are instead called \textit{Truths} and ReTweets are instead called \textit{ReTruths}. Note that throughout the remainder of this paper we most commonly refer to \textit{Truths} as \textit{posts} in order to avoid confusion with the epistemic use of truth (as in true/false). Due to the political and social circumstances surrounding its creation and launch, Truth Social has positioned itself as a hub for right-wing social media users disgruntled by mainstream platforms’ attempts to root out hateful and harmful communities and content.

Since its inception, and largely due to the influence of the former president's use of the platform, Truth Social has increasingly dominated a space of social media platforms that cater to users affiliated with the alt-right political movement---a technology space sometimes referred to as \textit{alt-tech} that also includes Parler, Gab, Rumble and others. Truth Social, in effect, functions as a kind of right-wing Twitter, but without the content regulation that is typically found in mainstream social media platforms. As a result, Truth Social, along with other alt-tech platforms, are a potential hotbed for misinformation, conspiracy theories, and other malign social media activity.

\begin{figure}[t]
\begin{center}
  \includegraphics[width=\linewidth]{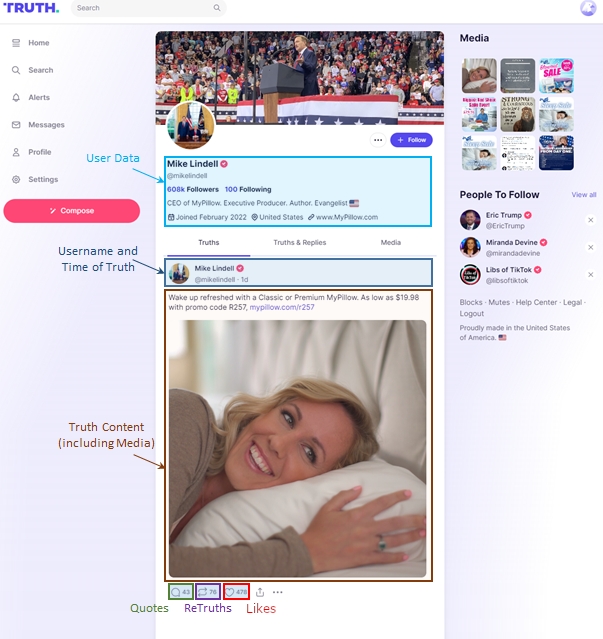}
  \caption{Annotated illustration of Truth Social Web Interface. The Web scraper extracted user data, and post information, including time, content (with links or other media), quotes, ReTruths, and likes.}
    
\end{center}
\end{figure}

Despite its scope and influence, little is known about the posts and content that is shared on the Truth Social network. The dearth of research involving Truth Social is partially due to how new it is, but also due to the lack of a publically available API. To ameliorate these issues, the current work presents a large dataset of Truths, ReTruths, users, and other data collected from a broad crawl over the Truth Social platform from its launch on February 21, 2022 until October 15, 2022. In total, this dataset contains the content of 823,927 Truths posted by 454,458 users including the full history of the 65,536 most active users. Truth Social does not publish its usage statistics, but we estimate that this dataset contains user data of about 20\% of the total registered users and an unknown, but larger, proportion of the total number of posts.

The dataset represents the first of its kind and can be used to ask and answer numerous research questions. For starters, social media's effects on people's consumption of information has become a topic of increasing importance~\cite{sharot2020people}. Providing users more direct agency over the information they consume, social media---while transformative---may limit exposure to diverse perspectives and cause the formation of like-minded users reinforcing shared narratives~\cite{cinelli2021echo}. This lack of exposure to diverse information and differing perspectives has been found to provide the scaffolding upon which conspiracy theories~\cite{cinelli2022conspiracy} and misinformation~\cite{del2016spreading} may grow. Truth Social is but the latest example of the formation of a self-referential, insular community---commonly called an echo chamber---that  is known to lead to increased political polarization \cite{barbera2020social}. Thus, because Truth Social is itself a right-wing echo chamber, catering to politically polarized defectors of mainstream social media, it provides a fertile ground for the spread of misinformation and conspiracy theories. Therefore, with the increased understanding of conspiracy theories' potentially damaging effects on democracy~\cite{sternisko2020dark}, understanding Truth Social's interaction with echo chambers and misinformation presents an important topic for continued research.

%%Patrick - I feel like these paragraphs repeat themselves, you can tell a singular story with these paragraphs?

% got it

% (notably Twitter~\cite{santucci_2021})
\subsection{A Brief Overview of Truth Social}
Truth Social's announcement and ultimate launch as a social media platform can be traced to former U.S. President Donald Trump's ban from several major social media platforms  following his role in the January 6 United States Capitol attack\footnote{\url{https://blog.twitter.com/en_us/topics/company/2020/suspension}}. As it currently stands, Truth Social is occupied largely by both users disaffected by mainstream platforms’ moderation policies and enthusiastic followers of Donald J. Trump, and appears generally similar to other ``alt-tech" platforms. However, Truth Social's status as the main mouthpiece for a former President whose influence remains momentous in the United States Republican Party positions it as a remarkably influential \textit{alt-tech} platform.

% and current de facto leader of the United States Republican Party positions it as a remarkably influential  ``alt-tech" platform.

% -- launched by a deplatformed former U.S. president and occupied largely by users disaffected by mainstream platforms’ moderation policies -- Truth Social is positioned as a uniquely influential ``alt-tech`` platform.

% POSSIBLE ALT SECOND PARAGRAPH
% Platforms with natures similar to Truth Social -- catering largely to users disaffected by the content moderation of mainstream social networks -- have been found to be decidedly successful in drawing users over from the original platform~\cite{papasavva2021qoincidence}, specifically in the case of followers of Donald Trump~\cite{horta2020does}. Moreover, these types of platform have been shown to both harbor and instigate dangerous conspiracy theories \cite{rye2020reading}. These conspiracy theories, while birthed on seemingly fringe platforms, have been found to ultimately jump to mainstream platforms~\cite{zannettou2017web}, thus advancing the same dangerous misinformation mainstream social networks attempted to thwart via de-platforming~\cite{tollefson2021tracking}.

Platforms with natures similar to Truth Social---catering largely to users disaffected by the content moderation of mainstream social networks---have been found to be decidedly successful in drawing users over from the original platform~\cite{papasavva2021qoincidence}, specifically in the case of followers of Donald Trump~\cite{horta2020does}. Moreover, these types of platforms have been shown to both harbor and instigate dangerous conspiracy theories~\cite{rye2020reading}, which, despite being birthed on seemingly fringe platforms, have been found to ultimately jump to mainstream platforms~\cite{zannettou2017web}, nevertheless advancing the same dangerous misinformation that mainstream social networks attempted to thwart in the first place~\cite{tollefson2021tracking}.

\section{Dataset Collection Methodology}
%Go over the methodology for how the data was collected here.
%Might want to include that the data is publicly available and we could not find anything prohibiting scraping
\label{sec:dataset_collection}
Collecting the posts and other activity data from Truth Social is not straightforward because the site does not provide a public API. Instead, we implemented a custom Web scraper and programmatically extracted the relevant context from Truth Social's Web interface directly. This Web interface did not impose any crawling restrictions nor did it disallow any crawling with the \texttt{robots.txt} standard. Because no API was available the most straightforward way to collect posts was from each specific account.

We crawled the Truth Social one account at a time, starting with @realDonaldTrump and then iteratively in a breadth first manner over the followers of each account. Specifically, the crawling methodology proceeded as follows:
\begin{enumerate}
  \item Collect information about the user's follower count, following count, creation date.
  \item Iterate through users following the user and users that the user follows, create an edge for each follower-followee relationship, and add that user to the breadth-first queue if that user has not been scraped in the past 14 days.
  \item Extract the full set of available Truths posted by the user.   
\end{enumerate}

This crawl began on September 4, 2022 and continued until October 14, 2022. In that time, all content posted by 65,536 users was collected. The dataset, therefore, has the complete set of all posts for the visited users before September of 2022.

One particular complication of the crawling methodology was the extraction of secondary-posts, \textit{i.e.}, ReTruths, Quotes, and Replies. During the initial crawl, we collected these secondary-posts but not the original post itself. So, at the end of the initial crawl, we additionally collected all of the original posts from which the initially collected ReTruths were based. Due to Truth Social's HTTP request limitations, we were only able to to connect the Quotes and Replies to the user to whom they were directed. Although these efforts resulted in a full accounting of the originating user account for all Quotes and Replies as well as the post content for all ReTruths, the opposite is not true. In other words, although we have the original user for all collected Quotes and Replies and the original post for all collected ReTruths, we we may not have collected all of the ReTruths, Quotes, or Replies for a given post. Nevertheless, as a result of our efforts the dataset is internally consistent.

\subsection{Data Model}

\begin{figure}[t]
    \centering
    \includegraphics[width=\linewidth]{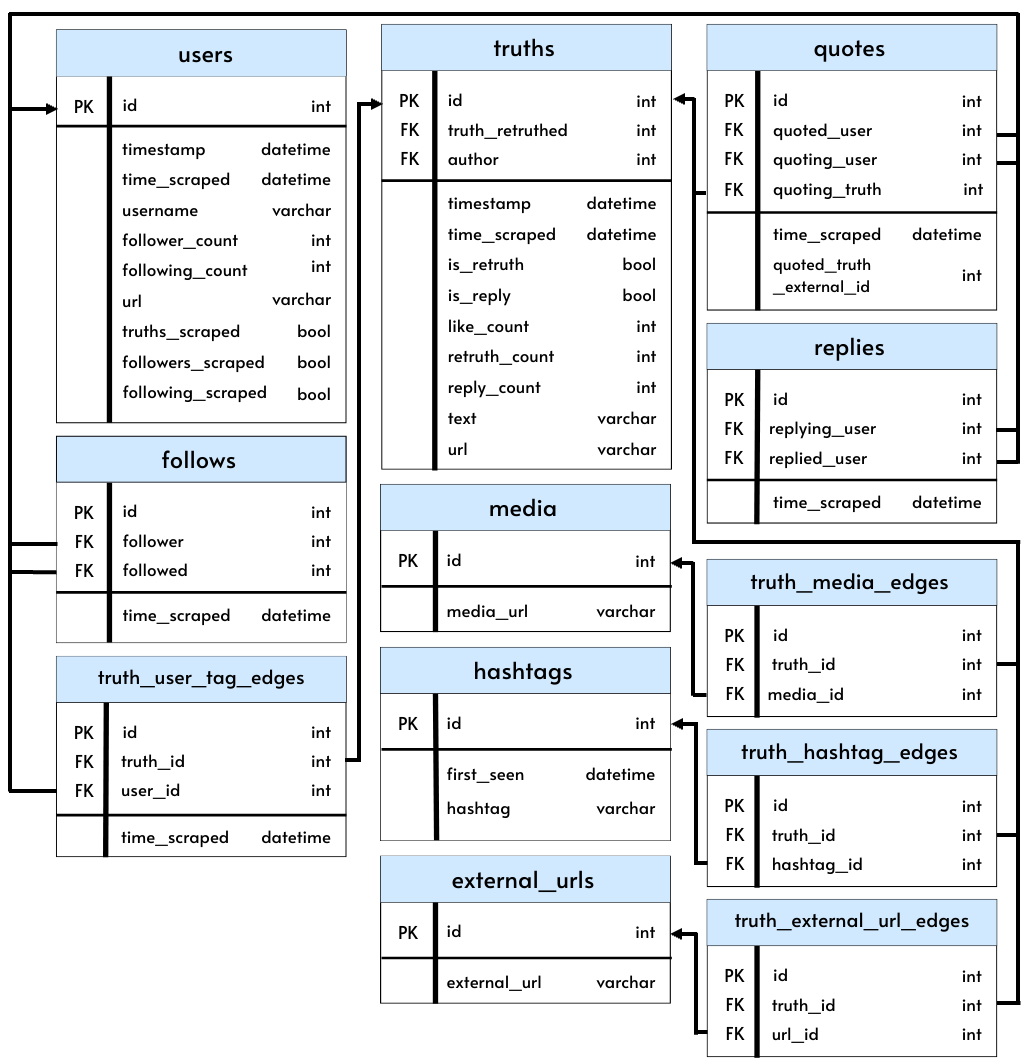}
    \caption{Relationship diagram between data elements in the collected schema. Arrows represent foreign key relationships among the tables.}
    \label{fig:db_diagram}
\end{figure}

During this crawl, data elements were stored in a local database system. The relational schema of this database is illustrated in Fig.~\ref{fig:db_diagram}. This dataset contains various kinds of related data and is therefore modeled as a relational database with foreign key dependencies. For example, follower/followee relationships are enumerated in the \texttt{follows} table, with each entry referencing the following and the followed user via a foreign key to the corresponding \texttt{user} record. Likewise, \texttt{truth} entries are linked to the corresponding \texttt{user}, and  quotes, replies, media, hashtags, external\_urls, and their respective edge tables are all linked with foreign key relationships to further contextualize the content the Truths.

\begin{table}[t]
\centering
\begin{tabular}{ll}
\toprule
Table           & Number of Records \\
\midrule
users.tsv           & 454,458       \\
follows.tsv         & 4,002,115       \\
truths.tsv        & 823,927       \\
quotes.tsv         & 10,508       \\
replies.tsv         & 506,276       \\
media.tsv         & 184,884       \\
hashtags.tsv & 21,599       \\
external\_urls.tsv          & 173,947       \\
truth\_hashtag\_edges.tsv      & 213,295       \\
truth\_media\_edges.tsv         & 257,500       \\
truth\_external\_url\_edges.tsv & 252,877       \\
truth\_user\_tag\_edges.tsv        & 145,234      \\
\end{tabular}
\caption{Description of the Truth Social Dataset}
\label{tab:datapoints}
\end{table}

Tables from the database system were exported to text files in a tab-delimited format and are available on the Zenodo data service at \url{https://doi.org/10.5281/zenodo.7531625} and a sample containing the first 10,000 records from each file are included as Supplemental Material. Table~\ref{tab:datapoints} lists the files available in the dataset and the number of records in each file.

\subsection{FAIR Principles}
The dataset presented by the present work conforms to the FAIR principles~\cite{wilkinson2016fair} and are therefore findable, accessible, interoperable, and re-usable:

\paragraph{Findable:} We provide the dataset publically using the Zenodo data service and give it a permanent digital object identifier (DOI) \url{https://doi.org/10.5281/zenodo.7531625}.
\paragraph{Accessible:} The dataset is freely available on the Internet and can be accessed by anyone with an internet connection. All of the data is provided as tab separated value (TSV) files, a standard format for handling data tabular data. 
\paragraph{Interoperable:} The dataset is easily loaded and viewed with most current database management systems or spreadsheet systems.
\paragraph{Re-usable:} Metadata is also included in a Readme file and is linked to the DOI of this paper for further reference.

\subsection{Limitations}
Although we endeavoured to capture a complete and holistic dataset, it is important to be aware of certain methodological and technological limitations and possible sampling biases that may be present in the dataset. Some of the key limitations are are follows:
\begin{enumerate}
    \item \textbf{Access to a user's followers is limited.} While Truth Social permits clients on its Web application to scroll through the entirety of a user's following list, it limits clients' access to a user's followers list to only 50 followers. It is unclear why this limitation exists and how these 50 followers are selected. It may be possible to estimate who follows whom by analyzing a complete following lists of all users and/or the users who frequently ReTruth, Quote or like another user. However, that analysis is not present in the current dataset.
    \item \textbf{Web Request Limits.} Although scraping limits are not published by Truth Social, we did endeavor to be responsible with the number of HTTP requests that were issued to Truth Social. This restricted our ability to capture more data.
    \item \textbf{Sampling Bias.} Recall that the crawling methodology proceeded in a breadth first manner from the a popular user @realDonaldTrump and then proceeded to other highly active users. As a result, this dataset likely contains the most active users of the platform. The choice of @realDonaldTrump as the seed-user may have also nudged the data collection towards more political users and posts. However, the average path length of the user network was relatively small and the post content is quite diverse, so we are confident that the sample is moderately representative of the whole platform.
\end{enumerate}
% \begin{itemize}
%     \item Cannot get all followers.
%     \item BFS structure -- may not have all connections from retruth to truth
%     \item Not all users are accompanied by their respective truths -- created user entries after the fact
% \end{itemize}

\subsection{Ethical Considerations}
The dataset is collected from a publically available resource. Users who submit content do so with the explicit intent of making their activity publically and widely visible. This research was observational only. No intervention or treatment was made to the population; therefore, this research was determined to be exempt from full ethical review panel by the University of Notre Dame institutional review board.

\section{Content Analysis}
To get a better understanding of the collected dataset we performed some rudimentary analysis on the content of the dataset. This analysis includes (1) an investigation into the top-linked Web sites, (2) a look at certain temporal artifacts of the text content of the posts, and (3) an analysis of the follower network of the platform.

\subsection{External Link Analysis}
Research has shown that misinformation campaigns often utilize and share external links to proliferate information across and between multiple platforms~\cite{wilson2020cross, golovchenko2020cross}. For this reason, we extracted the external Web links that users posted. We found 173,947 links in total. The domain of each link was extracted and aggregated for further analysis. 

\begin{figure}[t]
    \centering
    % This file was created with tikzplotlib v0.10.1.
\begin{tikzpicture}

\definecolor{darkgray176}{RGB}{176,176,176}

\begin{axis}[
xlabel={Linked Domain Frequency in Truths},
ylabel={Linked Domain Frequency in ReTruths},
xmode=log, ymode=log,
y grid style={darkgray176},
ymin=60, xmin=60,
ymax=10000000, xmax=1000000,
]
\addplot [
  mark=*,
  only marks,
  scatter,
  nodes near coords,
  scatter/@post marker code/.code={%
  \endscope
},
  scatter/@pre marker code/.code={%
  \expanded{%
  \noexpand\definecolor{thispointdrawcolor}{RGB}{\drawcolor}%
  \noexpand\definecolor{thispointfillcolor}{RGB}{\fillcolor}%
  }%
  \scope[draw=thispointdrawcolor, fill=thispointfillcolor]%
},
  visualization depends on={value \thisrow{draw} \as \drawcolor},
  visualization depends on={value \thisrow{fill} \as \fillcolor}
]
table{%
x  y  draw  fill
67259 3106672 191.25,0,191.25 191.25,0,191.25
9839 698284 191.25,0,191.25 191.25,0,191.25
2750 500040 0,0,255 0,0,255
2272 450811 0,0,255 0,0,255
4274 421838 0,0,255 0,0,255
3297 389997 0,0,255 0,0,255
6244 387704 191.25,0,191.25 191.25,0,191.25
1903 331972 0,0,255 0,0,255
2342 238575 0,0,255 0,0,255
1483 214983 0,0,255 0,0,255
11035 115239 255,0,0 255,0,0
6669 4054 255,0,0 255,0,0
6054 157 255,0,0 255,0,0
5277 13348 255,0,0 255,0,0
5007 129786 255,0,0 255,0,0
4880 108 255,0,0 255,0,0
4837 208491 255,0,0 255,0,0
};
\draw (axis cs:67259,3106672) node[
  font=\sffamily\tiny,
  anchor=east,
  text=black,
  rotate=0.0
]{rumble};
\draw (axis cs:9839,698284) node[
  font=\sffamily\tiny,
  anchor=west,
  text=black,
  rotate=0.0
]{justthenews};
\draw (axis cs:2750,500040) node[
  font=\sffamily\tiny,
  anchor=south east,
  text=black,
  rotate=0.0
]{babylonbee};
\draw (axis cs:2272,450811) node[
  font=\sffamily\tiny,
  anchor=east,
  text=black,
  rotate=0.0
]{conservativebrief};
\draw (axis cs:4274,421838) node[
  font=\sffamily\tiny,
  anchor=south,
  text=black,
  rotate=0.0
]{oann};
\draw (axis cs:3297,389997) node[
  font=\sffamily\tiny,
  anchor=north,
  text=black,
  rotate=0.0
]{hannity};
\draw (axis cs:6244,387704) node[
  font=\sffamily\tiny,
  anchor=west,
  text=black,
  rotate=0.0
]{breitbart};
\draw (axis cs:1903,331972) node[
  font=\sffamily\tiny,
  anchor=east,
  text=black,
  rotate=0.0
]{rsbnetwork};
\draw (axis cs:2342,238575) node[
  font=\sffamily\tiny,
  anchor=north,
  text=black,
  rotate=0.0
]{thefederalist};
\draw (axis cs:1483,214983) node[
  font=\sffamily\tiny,
  anchor=east,
  text=black,
  rotate=0.0
]{dcenquirer};
\draw (axis cs:11035,115239) node[
  font=\sffamily\tiny,
  anchor=west,
  text=black,
  rotate=0.0
]{thegatewaypundit};
\draw (axis cs:6669,4054) node[
  font=\sffamily\tiny,
  anchor=west,
  text=black,
  rotate=0.0
]{foxsports};
\draw (axis cs:6054,157) node[
  font=\sffamily\tiny,
  anchor=west,
  text=black,
  rotate=0.0
]{gamespot};
\draw (axis cs:5277,13348) node[
  font=\sffamily\tiny,
  anchor=west,
  text=black,
  rotate=0.0
]{telegram};
\draw (axis cs:5007,129786) node[
  font=\sffamily\tiny,
  anchor=125,
  text=black,
  rotate=0.0
]{youtube};
\draw (axis cs:4880,108) node[
  font=\sffamily\tiny,
  anchor=west,
  text=black,
  rotate=0.0
]{hiphopdx};
\draw (axis cs:4837,208491) node[
  font=\sffamily\tiny,
  anchor=west,
  text=black,
  rotate=0.0
]{foxnews};
\end{axis}

\end{tikzpicture}
    \caption{Top 10 linked domains appearing in Truths (red, x-axis) and top 10 linked domains appearing in ReTruths (blue, y-axis). Purple marks indicate that the domain is in the top 10 in both Truths and ReTruths.}
    \label{fig:external_links}
\end{figure}
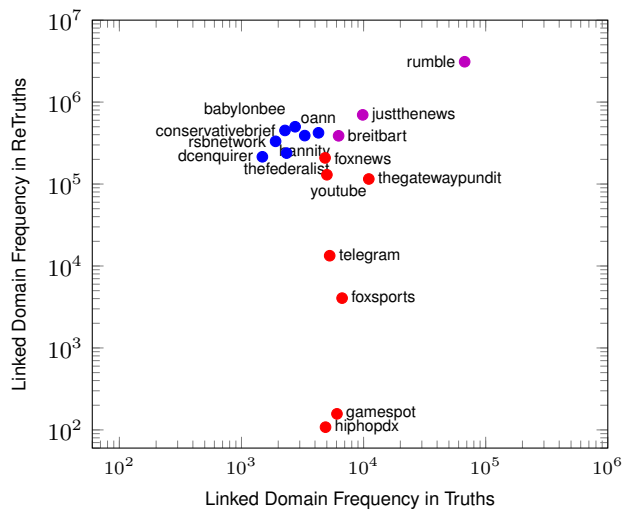

Figure~\ref{fig:external_links} illustrates the top linked domains in terms of the number of Truths in which they appear (x-axis) and the number of ReTruths in which they appear.

% Website with great info on Rumble. https://imge.com/what-is-rumble/
The top linked external Web site in terms of frequency in both Truths and ReTruths is Rumble, a video sharing platform. This site is known for having looser content moderation restrictions compared to many other video sharing platforms like YouTube and has become a haven for controversial figures that have been banned from mainstream platforms. Rumble has grown in popularity in recent years at least partially due to its use by former President Donald Trump to stream his political rallies and post news clips.

Other popular site include right-wing political networks like OANN, Breitbart, FoxNews, etc. Curiously, a large number of external links point to the Telegram social media platform. Popular in Russia, Telegram is frequently used by supporters of Russia in the Russia-Ukraine War~\cite{theisen2023motif}. It has also in been investigated as a platform that acts as a safe haven for those that have been deplatformed from other social media sites~\cite{rogers2020deplatforming}. Because Telegram has this association with deplatformed individuals we further investigated the five most shared telegram channels to better understand the type of content that was shared. In order of decreasing popularity, the five most shared telegram channels are:
\begin{enumerate}
    \item \textbf{RealKarliBonne}. This channel shares a large variety of memes, Truths, and messages supporting Donald Trump. The channel also shares a large amount of video content featuring the Biden administration. Although some video messages are supportive of Trump, the preponderance is negative of the Biden administration and may serve as an example of affective polarization~\cite{iyengar2019origins}.
    \item \textbf{LauraAbolichannel}. This channel does not appear to have a clear focus. Its content contains commentary and clips from right-wing media, anti-vaccination messages, and uplifting memes.
    \item \textbf{freedomforcebattalion}. This channel presents former President Donald Trump's messaging and other current events with a biblical perspective. 
    \item \textbf{realx22report}. This is the official Telegram channel of X22report, a daily show that covers financial and political issues. 
    \item \textbf{drawandstrikechannel} This is the channel of right-wing political columnist Brian Cates. Cates is an active user on Truth Social and draws a major following as well in his Telegram channel. Content in the channel focuses around various political topics and current events.
\end{enumerate}

The combination of Truth Social's politically charged user-base and external links' role in narrative-building and misinformation campaigns creates a significant space for further analysis in both the information contained in the external links and the temporal aspect of the external links.

Examinations into the information contained in the external links may provide major insights into the spread of conspiracy theories and misinformation on Truth Social and across the social Web. For example, to better understand the popularization of conspiracy theories on Truth Social, one may trace the earliest external domains that references a certain conspiracy theory and analyze their role in the narrative spread through Truth Social.

%Examinations into the temporal aspect of external links, \textit{i.e.}, which domains dominate at what point, may provide major insights into the external influences with the greatest effect on Truth Social's users. For example, external links may play a major role in building cohesive narratives around certain events, and further analysis into how narratives jump from external domains to and from Truth Social users is needed to understand how external agents may influence social media users broadly.

Overall, external links in Truth Social remarkably reflect the politically charged user-base, and therefore external agents' influence on the user-base cannot be overlooked. Thus, we believe that further analysis of these external links' role in the Truth Social network is critical to understanding how certain narratives and conspiracy theories propagate across the network.

\begin{figure*}
    \centering
    \input{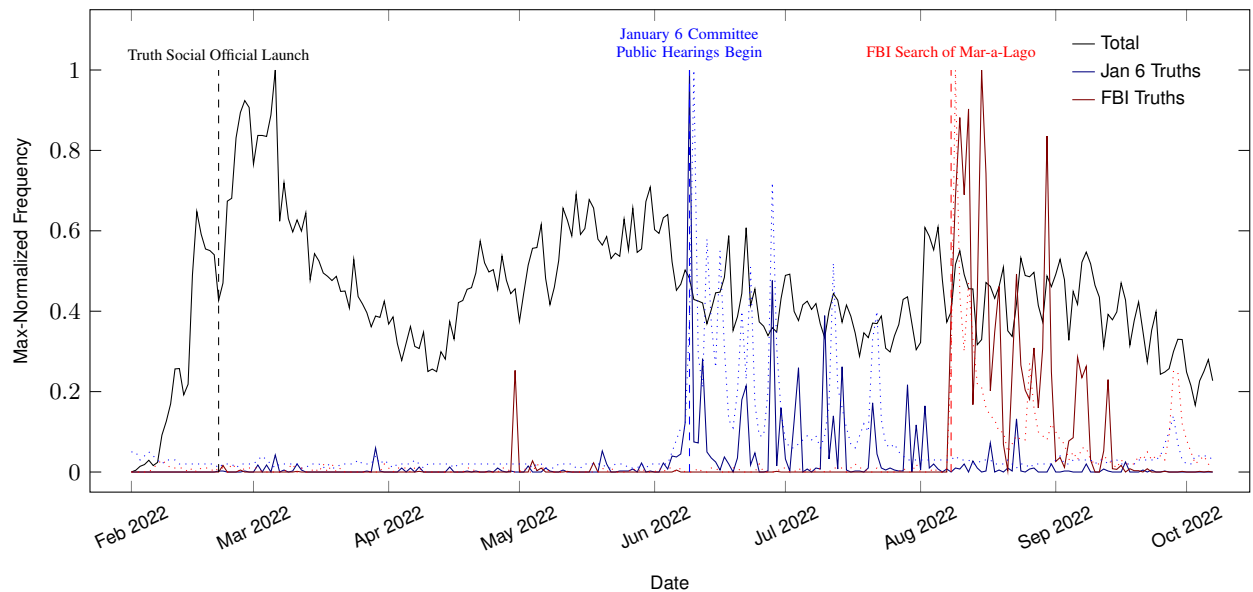}
    \caption{Daily max-normalized frequency of Truths posted in the dataset. Because Truth Social has a strong affiliation with former United States President Donald Trump, political activity and news stories involving the former president are commonly discussed on the site. In this figure, posts mentioning political events like the US House of Representative's Special Committee on the January 6 riots at the US Capitol (blue) and the FBI search at the former president's residence (Mar-a-Lago) spike surrounding these events. Dotted blue and red lines illustrate the data from Google Trends corresponding to the event.}
    \label{fig:daily_truth}
\end{figure*}

\subsection{Text Analysis}
To better understand both the temporal nature of the dataset and the possible interaction between external events and popular topics on Truth Social, we next performed a text analysis on the 823,927 posts in the dataset.

Figure~\ref{fig:daily_truth} provides an example illustration of our findings. Max-normalized by posts per day, the frequency of all posts provides a backdrop for our further evaluation into posts containing certain keywords. Overall, we found that the daily frequency of posts was highest near the official launch of Truth Social, before falling slightly and stabilizing.

Because of its affiliation with the former president, posts on Truth Social largely revolve around activities related to conservative politics and news stories. To illustrate this more concretely we considered two events that occurred during dates covered by the data collection methodology involving the former president: (1) the public hearings from the United States' House of Representative's Select Committee on the January 6 riots at the Capitol Building, and (2) the FBI raid of the former president's residence commonly called Mar-a-Lago.

Our first text analysis centers on the January 6 United States Capitol attack. Utilizing the keywords ``January 6'', ``January Six'', ``Jan 6'', and ``Jan Six'', we identified posts containing any of these phrases (with case-insensitive criteria). As shown in solid blue in Fig.~\ref{fig:daily_truth}, immediately following the start of the January 6 Committee's Public Hearings, posts containing these phrases spiked and continued to be elevated for approximately ten weeks. We believe that further evaluation may find correlation between these repetitive spikes and the ensuing broadcasts of the January 6 Committee's public hearings, but a thorough analysis is outside the scope of the current paper.

Next, we analyzed posts related to the FBI raid of Mar-a-Lago, utilizing the keywords ``Mar-a-Lago'' and ``Mar a Lago''. Illustrated in solid red in Fig.~\ref{fig:daily_truth}, we again identified posts containing any of these phrases (with case-insensitive criteria). Like in the January 6 example above, in this case we also found that Truths containing these phrases spiked following the events.

Dotted blue and red lines in Fig.~\ref{fig:daily_truth} illustrate the data from Google Trends for the January 6 and Mar-a-Lago searches, respectively. We see that the attention of these two events are symmetrical across Truth Social and the Web in general. This symmetry shows that the topics discussed on Truth Social appears to be largely representative of activity on the Web. 

When evaluating these spikes, we found that they were often driven largely by a small number of original posts that were ReTruthed many times. Some of the most popular posts originated from the former president himself, 
but were elevated and expanded by popular commentators, whose roles in narrative-building on Truth Social is a clear topic for further inspection. For example, following the FBI search of Mar-a-Lago, the former president posted ``A horrible thing that took place yesterday at Mar-a-Lago. We are no better than a third world country, a banana republic. It is a continuation of Russia, Russia, Russia, Impeachment Hoax \#1, Impeachment Hoax \# 2, the no collusion Mueller Report, and more. To make matters worse it is all, in my opinion, a coordinated attack with Radical Left Democrat state and local D.A.'s and A.G.'s''. This set off a conspiracy-laden narrative of the events that echoed throughout the platform. These posts typically gathered tens of thousands of ReTruths and likes---a large number for the relatively small platform. These findings and the dataset as a whole may provide much needed insight into the cohesion of the Truth Social platform as well as a better understanding of the role that popular commentators play in driving the narrative for the majority of users.

Ultimately, external events appear to have significant influence on Truth Social's user network. Moreover, the near-immediate rise in posts following certain events may point to both the interests of users and the cohesiveness of the Truth Social network. Further work lies in both the examination of narratives and sentiments accompanying apparent reactions to external events, as well as the examination of the platform-wide influence of a small group of popular commentators and if the narratives and stories that percolate within Truth Social eventually make their way onto mainstream social media platforms.

\subsection{Network Analysis}
Like Facebook, Instagram, and Twitter, the friendship network (via followers or ReTruths) of the Truth Social platform forms a social network that can be analyzed to find social roles, and network-based artifacts like centrality, betweeness, cliques, and other interesting network-based phenomena.

Because the follower graph could not be reliably extracted from the Web interface, we instead used ReTruths to construct the network. Although not the same as the follower graph, ReTweets (the Twitter-analog to ReTruths) are known to ``more closely mirror real-world relationships and trust''~\cite{bild2015aggregate}. 

\begin{figure}
    \centering
    \input{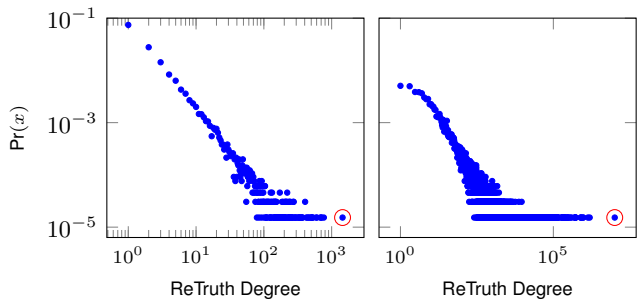}
    \caption{Degree distribution of the ReTruth graph by distinct users (left) and by total number of ReTruths (right). The red circle represents @realDonaldTrump who has the most ReTruths according to both criteria.}
    \label{fig:graph}
\end{figure}

We used two different criteria to analyze the network. First, we found the number of unique, fully-scraped users who ReTruthed a user. Figure~\ref{fig:graph} (left) illustrates the degree distribution of the ReTruth graph, that is, the number of distinct users who Retruthed another user. This distribution appeared to follow the generally followed typical power-law degree distribution found on Twitter. The user with the highest number of distinct ReTruthing users was ReTruthed by 1,448 users.

Rather than looking at distinct users, our next analysis looked at the total number of ReTruths for each user. Figure~\ref{fig:graph} (right) shows that the ReTruth multi-graph, where multiple ReTruths from the same user count as multiple edges, has a slightly different degree distribution that more closely resembles a log-normal distribution. The user with the highest number of  ReTruths was ReTruthed 1,0176,833 times.

\section{Conclusion}
This paper presents a large dataset, which contains information of 454,000 users and over 823,000 posts including the complete history of the 65,536 most active users from the Truth Social platform. In addition, this dataset covers the ReTruths, quotes, text, media, and other information from the platform. We also perform a preliminary analysis of this dataset. 

Our preliminary analysis shows that a handful of external Web sites dominate Truth Social posts, with Rumble appearing most frequently. Moreover, a brief look at the most commonly linked Telegram channels finds that right-wing and Trump-focused Telegram channels appear most often. We also analyzed the temporal and structural nature of the posts and ReTruths in the dataset. We have uncovered several interesting findings and avenues for further study.

% number of Truths posted containing certain keywords, we were able
% Additionally, our analysis points to a relatively constant number of Truths posted per day, as well as a

% Additionally, our analysis illustrates how external events may have influenced behavior on Truth Social, specifically the U.S. January 6 Committee~\cite{broadwater_haberman} and the Mar-a-Lago Search. Finally, our analysis shows that posts on Truth Social have remained relatively constant after an initial spike at its launch.

Overall, Truth Social is an emerging alt-tech platform. Targeted at hard-right users disaffected from mainstream social media platforms and working as the main mouthpiece of former President Donald Trump, Truth Social's unique position in the information ecosystem cannot be overlooked. This dataset provides researchers a means to study the Truth Social platform, permitting research on both Truth Social itself and important socio-technical issues including the cultivation and spread of information and narratives.

\bibliography{citations}

\end{document}